\documentclass[superscriptaddress,prc,twocolumn,nofootinbib]{revtex4}

\usepackage{graphicx}
\usepackage{amssymb,amsmath}
\usepackage[utf8]{inputenc}
\usepackage{multirow}
\usepackage{xcolor}
\usepackage{hyperref}
\hypersetup{colorlinks=true,linkcolor=blue,anchorcolor=blue,citecolor=blue,filecolor=blue,urlcolor=blue,bookmarksnumbered=true}
\usepackage[flushleft]{threeparttable}

\usepackage{amsmath}% http://ctan.org/pkg/amsmath

\begin{document}
\title{Calculations of \boldmath$p(n,\gamma )d$ reaction in chiral effective field theory}
\author{Weijie Du}
\email[Email:]{ duweigy@gmail.com}
\affiliation{Department of Physics and Astronomy, Iowa State University, Ames, Iowa 50010, USA}
\author{Soham Pal}
\affiliation{Department of Physics and Astronomy, Iowa State University, Ames, Iowa 50010, USA}
\author{Mamoon Sharaf}
\affiliation{Department of Physics and Astronomy, Iowa State University, Ames, Iowa 50010, USA}
\author{Peng Yin}
\affiliation{Department of Physics and Astronomy, Iowa State University, Ames, Iowa 50010, USA}
\affiliation{Institute of Modern Physics, Chinese Academy of Sciences, Lanzhou 730000, China}
\author{Shiplu Sarker}
\affiliation{Department of Physics and Astronomy, Iowa State University, Ames, Iowa 50010, USA} 
\author{Andrey M. Shirokov}
\affiliation{Skobeltsyn Institute of Nuclear Physics, Lomonosov Moscow State University, Moscow 119991, Russia} 
\author{James P. Vary}
\affiliation{Department of Physics and Astronomy, Iowa State University, Ames, Iowa 50010, USA}
\date{\today}
\begin{abstract}
We present a calculation of the radiative capture cross section $p(n,\gamma )d$ in the low-energy range, where the $M1$ reaction channel dominates. 
Employing the LENPIC nucleon-nucleon interaction up to the fifth order (N4LO) that is regularized by the semi-local coordinate space regulators, we obtain the initial and final state wave functions, and evaluate the phase shifts of the scattering state and deuteron properties.
We derive the transition operator from the chiral effective field theory up to the next-to-next-to leading order (N2LO), where we also regularize the transition operator using regulators consistent with those of the interactions. 
We compute the capture cross sections and the results show a converging pattern with the chiral-order expansion of the nucleon-nucleon interaction, where the regulator dependence of the results is weak when higher-order nucleon-nucleon interactions are employed. 
We quantify the uncertainties of the cross-section results due to the chiral-order truncation.
The chirally complete and consistent cross-section results are performed up to N2LO and they compare well with the experiments and other theoretical predictions. 
\end{abstract}
\maketitle

\section{Introduction}

Nuclear physics plays a fundamental role in studying the evolution of the universe \cite{Thompson:2009,Phillips:2016mov}. Nuclear astrophysics is nowadays an open field and requires accurate input from nuclear physics \cite{Burles:1999zt,Burles:2000zk}. However, direct measurements of the cross sections at stellar energies are challenging as many relevant cross sections occur in the experimentally challenging low-energy range \cite{Phillips:2016mov,Thompson:2009}. It is thus important to develop advanced experimental techniques \cite{Broggini:2010}. Meanwhile, it is equally important to develop first-principles microscopic theories with predictive power.

A promising theoretical approach is the chiral effective field theory ($\chi$EFT) \cite{Weinberg:1991um,Weinberg:1990rz,Scherer:2012book} in combination with model-independent {\it ab initio} few- and many-body methods (see Ref.~\cite{Maris:2020qne} and references therein). The $\chi$EFT describes the nuclear interactions \cite{Entem:2003ft,Machleidt:2011zz,Epelbaum:2008ga,Epelbaum:2014sza,Epelbaum:2014efa} based on the underlying fundamental theory (quantum chromodynamics). It is also employed to derive single-, two-, and multi-nucleon electroweak currents \cite{Pastore:2008ui,Kolling:2009iq,Pastore:2009is,Pastore:2011ip,Piarulli:2012bn,Schiavilla:2018udt,Baroni:2018fdn,King:2020wmp,Epelbaum:2017NTSE,Kolling:2011mt,Krebs:2019aka,Krebs:2020}. The model-independent {\it ab initio} methods utilize direct input from the $\chi$EFT, where the consistent scheme of the power expansion for both the nuclear interactions and the nucleon currents enables systematic and quantified convergence study and uncertainty analysis of the calculations \cite{Maris:2020qne,Epelbaum:2014sza,Epelbaum:2014efa,Epelbaum:2008ga}. 

In this prototypical study, we aim to perform first-principles calculations for nuclear reactions in the energy range of astrophysical interest. For the demonstration purpose, we study the radiative capture process $p(n,\gamma )d$, which is one of the simplest reactions, yet playing a critical role in big bang nucleosynthesis \cite{Thompson:2009,Pisanti:2007hk,Cooke:2017cwo}. The experimental data for this reaction is sparse in the low energy range \cite{Cox:1965,Cokinos:1977zz,Tomyo:2003qyn,Nagai:1997zz}. Therefore, theoretical studies with predictive power are especially needed. 

Indeed, various predictive methods have been developed for precision calculations of the reaction $p(n,\gamma )d$ based on either the pionless effective field theory \cite{Chen:1999bg,Rupak:1999rk,Beane:2000fi,Ando:2006,Beane:2015yha} or the pionful $\chi$EFT \cite{Piarulli:2012bn,Acharya:2022} as alternatives to traditional approaches based on phenomenological models for the nuclear interactions and currents \cite{Arenhovel:1990yg,Carlson:1997qn,Marcucci:2004sq}. In view of these successful advances, it becomes important to explore new approaches to reaction calculations that can be used in combination with microscopic {\it ab initio} methods to study many-body nuclear bound states like the no-core shell model (NCSM)~\cite{Barrett:2013nh,Navratil:2000gs,Navratil:2000ww} and resonances like the SS-HORSE-NCSM approach~\cite{Shirokov:2016thl,Shirokov:2016ywq,Blokhintsev:2017lxs,Blokhintsev:2017rae,PRC239,YaF245}. Knowledge along this research line can be generalized to treat complicated reactions that include complex nuclei.

In this work, we focus on the low-energy range on $p(n,\gamma )d$ radiative capture where the $M1$ transition channel dominates. We construct the Hamiltonian matrices for the initial and final nuclear systems with the LENPIC $NN$ interaction~\cite{Epelbaum:2014sza,Epelbaum:2014efa} that are derived from $\chi$EFT. We compute the bound and scattering state wave functions via the direct matrix diagonalization and the method of harmonic oscillator representation of scattering equations (HORSE) \cite{Bang:2000bang}, respectively. We also develop the chiral $M1$ transition operators \cite{Kolling:2009iq,Kolling:2011mt,Krebs:2019aka} with the same semi-local coordinate space (SCS) regulators as those adopted for the LENPIC $NN$ interaction. We perform calculations of the capture cross section and quantify the uncertainty of the results due to the order-by-order truncation of the $NN$ interaction. Chirally consistent and complete calculations are achieved up to the next-to-next-to leading order (N2LO). Our work complements the work by Piarulli et al. \cite{Piarulli:2012bn} and that by Acharya and Bacca \cite{Acharya:2022}. However, the difference in the microscopic approaches, the input $NN$ interactions, the choice of regulator schemes, and the systematic convergence and uncertainty analyses distinguish our work.  

In Sec. \ref{sec:TheoryPart}, we present the elements of the theory, which include the details of computing the initial and final state wave functions, our transition operators, and the cross section. In Sec. \ref{sec:ResultAndDiscussionPart}, we show the results of the observables of the deuteron (final state), the phase shifts of the scattering waves (initial state), and the capture cross sections. We conclude in Sec. \ref{sec:SummaryAndOutlook}, where we also provide an outlook.

\section{Theory}
\label{sec:TheoryPart}

In this work, we compute the radiative capture cross section of the $p(n,\gamma )d$ reaction at low center-of-mass (CM) bombarding energy ($\leq 0.01$ MeV) in the relative coordinates of the neutron-proton ($np$) system. The $M1$ reaction dominates \cite{Blatt:2010}\footnote{The contribution from the competing electric dipole $E1$ reaction channel increases with energy. However, even at the highest CM bombarding energy 
examined here (0.01 MeV), the contribution from the $E1$ reaction channel is about 1$\%$ of the total capture cross section \cite{Rupak:1999rk}.}: the initial (scattering) state of the $np$ system is in the ${^1}S_0$ channel, while the final (bound) state is in the $^{3}SD_{1}$ channel; and a photon is emitted with the excess energy of the nuclear system during the reaction.

The capture cross section can be calculated based on the transition matrix element \cite{Blatt:2010}
\begin{align}
\mathcal{M}_{\lambda} (i, f) = \langle \phi_{f} | { \mu }_{1 \lambda} | \phi_{i} \rangle , \label{eq:dipole-transition-integral}
\end{align}  
where ${ \mu }_{1 \lambda }$ ($\lambda =0, \pm 1$) denotes the $\lambda$-component of the $M1$ transition operator $ \vec{ \mu } $ (rank-one tensor). $| \phi_{i} \rangle $ and $ | \phi_{f} \rangle $ denote the initial and final state vectors of the $np$ system, respectively. 

Following the discussion in Ref. \cite{Blatt:2010}, we start with the scattering setup where both projections of the total angular momenta of $| \phi_{i} \rangle $ and $ | \phi_{f} \rangle $ are zeroes, without the loss of generality.\footnote{One notes that this transition probability is the same as those computed with the other two choices of the final state polarization, i.\:e., the projection of the total angular momentum of $ | \phi_{f} \rangle $ being $+1 (-1)$, where compatible transition operator ${ \mu }_{1-1}$(${ \mu }_{1+1}$) should be adopted. This can be seen by applying the Wigner--Eckart theorem \cite{Suhonen:2007} to Eq. \eqref{eq:dipole-transition-integral}.} In this case, only the $\lambda =0$ component of the $M1$ transition operator contributes. The transition probability can be written as \cite{Blatt:2010}
\begin{align}
T_M(i, f) = \frac{16 \pi}{9} \kappa ^3 |\mathcal{M}_0 (i, f)|^2 , \label{eq:Transition_Probability_Equivalence}
\end{align}
where $\kappa $ is the wave number of the emitted photon. We adopt the natural units and set ${\hbar=c=1}$ in this work.

Since there are three possible transitions from the ${^1}S_0$ state to the deuteron state (with different polarizations) and these transitions are of equal probability, the total $M1$ radiative capture cross section for the unpolarized $np$ system to form a deuteron can be calculated as \cite{Blatt:2010}
\begin{align}
\sigma  = \frac{3}{4 \Phi  } T_M(i, f) , \label{eq:CaptureCrossSection}
\end{align}
where one sums over the possible polarizations of the final state and averages over the initial polarizations of the $np$ system. $\Phi $ denotes the flux of the scattering wave. 

In the remainder of this section we describe the methods for solving the Schr\"odinger equation and calculating the initial and final state wave functions of the reaction. We also derive the $M1$ transition operator based on the $\chi$EFT. Though the current work focuses on the two-body reaction problem, our methodology can be generalized to study similar reactions using wave functions from {\em ab initio} many-body NCSM calculations \cite{Barrett:2013nh,Navratil:2000gs,Navratil:2000ww,Blokhintsev:2017lxs,Blokhintsev:2017rae,Shirokov:2016thl,Shirokov:2016ywq,PRC239,YaF245}.

\subsection{Nuclear Hamiltonian and basis representation}
The Hamiltonian of the $np$ system in relative coordinates is
\begin{align}
H = T_{\rm rel} + V_{\rm NN},
\label{eq:Hd-H}
\end{align}
where $T_{\rm rel}$ denotes the relative (intrinsic) kinetic energy and $V_{\rm NN}$ the inter-nucleon interaction. 

We employ the three-dimensional harmonic oscillator (3DHO) basis to construct the matrix representation of the Hamiltonian and various operators throughout this work. The wave functions are expanded in a series of 3DHO basis functions that is useful for a straightforward generalization of our approach to studies of many-body nuclear systems within NCSM~\cite{Barrett:2013nh,Navratil:2000gs,Navratil:2000ww} or other {\em ab initio} approaches.

The 3DHO basis functions of relative motion are specified as $|nlSJM_J \rangle $, where $n$ is the radial quantum number, $l$ is the orbital angular momentum, and $S$ is the total spin of the $np$ system. The total angular momentum $J$ is coupled from $l$ and $S$, whereas $M_J$ denotes the projection of $J$. The oscillator quanta is $2n+l$, which is an index to scale the basis dimension. We remark that $S$, $J$, $M_J$ and the parity $P_{\pi}=(-1)^l$ are the good quantum numbers specifying the $np$ system. 

In the coordinate representation, the 3DHO basis reads
\begin{multline}
 \langle \vec{r} | nlSJM_J \rangle\\
 =R_{nl} (r) \sum _{m_l,m_{s}} 
 (lm_lSm_s|JM_J) Y_{lm_l}(\Omega _{\hat{r}}) \chi _{Sm_s} 
\label{eq:3DHO_r_rep},
\end{multline}
where the summations over $m_l$ and $m_s$ run over all the possible values of the projections of $l$ and $S$, respectively. $(lm_lSm_s|JM_J)$ denotes the Clebsch--Gordan coefficient and $Y_{lm_l}(\Omega _{\hat{r}}) $ is the spherical harmonics (we adopt the Condon--Shortly convention \cite{Suhonen:2007} in this work). $\chi _{Sm_s} $ is the spin part of the wave function. The radial part of the oscillator functions is
\begin{align}
R_{nl}(r)=\sqrt{\frac{2n!}{r_0^3 \Gamma (n+l + \frac{3}{2}) }} \Big( \frac{r}{r_0} \Big)^l e^{-\frac{r^2}{2r_0^2}} L_n^{l+\frac{1}{2}} \Big( \frac{r^2}{r^2_0} \Big)
\end{align}
with  $L_n^{l+\frac{1}{2}} \big( {r^2}/{r^2_0} \big)$ and $ \Gamma (n+l + {3}/{2}) $ 
being respectively the associated Laguerre polynomial and the Gamma function~\cite{Arfken:2011}. The characteristic length scale of the 3DHO basis can be expressed as $r_0=(\widetilde{m}_N \omega )^{-1/2}$, where $\omega$ denotes the oscillator energy and $\widetilde{m}_N $ is the reduced mass of the neutron and proton (set to be $469.46$ MeV in this work). 

\subsection{The initial state}
For our application to very low incident energy, we restrict our discussion of the $np$ scattering to the uncoupled channel that is specified by the quantum numbers $l$, $S$, $J$, and $M_J$. We calculate the initial state of the $np$ system via the HORSE method \cite{Bang:2000bang, Yamani:1975, Zaytsev:1998,Shirokov:2003kk}. Based on Eq. (1) in Ref. \cite{Bang:2000bang} (where the spin-part of the scattering wave function is ignored), we construct the incoming scattering wave function using the partial wave expansion and couple the orbital angular momentum to the spin of the $np$ system. Taking the relative momentum of the initial $np$ system $\vec{k}$ to be along the $\hat{z}$ axis, the scattering wave function in the uncoupled channel is
\begin{multline}
 \langle \vec{r} | \phi _i (\vec{k}) \rangle %\\
 = \frac{1}{k} \sum _{n=0}^{\infty}  \sqrt{4\pi (2l+1)}\; a_{nl}(k)  R_{nl}(r) \\
  \times \sum _{m_s} (l 0 S m_s|J M_J) Y_{l0}(\theta , \varphi ) \chi _{Sm_s}  , 
   \label{eq:scatteringWaveFunction}
\end{multline}
where $\vec{r}$ determines the relative position of the nucleons with $r=|\vec{r}|$ and $k=|\vec{k}|$. The polar angle $\theta $ is defined as $\cos \theta = \vec{k}\cdot \vec{r}/(kr) $. $\varphi $ denotes the azimuthal angle. The amplitudes of the 3DHO basis expansion of the wave function are $\{ a_{nl}(k) \} $. We normalize the scattering wave function such that the flux $\Phi $ associated with the scattering wave is unity.

The scattering state of the $np$ system satisfies the Schr\"odinger equation
\begin{align}
H | \phi _i (\vec{k}) \rangle = E | \phi _i (\vec{k}) \rangle ,
\end{align}
where $E = {k^2}/{(2\widetilde{m}_N)}$ is the energy in the CM frame. In the 3DHO basis, the Schr\"odinger equation is equivalent to the following set of the algebraic equations
\begin{multline}
\sum _{n'=0}^{\infty} \  \langle nlSJM_J |  H - E \delta _{nn'}  | n'lSJM_J \rangle \\
\times \langle n'lSJM_J | \phi _i (\vec{k}) \rangle  = 0 ,
 \label{eq:scatteringSystem}
\end{multline}
from which we can obtain the amplitudes $a_{nl}(k) = \langle nlSJM_J | \phi _i (\vec{k}) \rangle $ for given values of $l$, $S$, $J$ and $M_{J}$.

The HORSE method solves Eq. \eqref{eq:scatteringSystem} for the amplitudes $a_{nl}(k)$ by truncating the interaction matrix element $ \langle nlSJM_J |  V_{\rm NN}  | n'lSJM_J \rangle$ in the Hamiltonian up to some large but finite dimension. In particular, one notes that the interaction matrix element decreases with increasing $n $ and $n'$, while the matrix element of the kinetic energy increases linearly with $n $ and $ n'  \rightarrow \infty $ \cite{Bang:2000bang,Shirokov:2016thl}. Therefore, a cutoff scale $\widetilde{n}$ is introduced to the interaction matrix in the Hamiltonian of the initial $np$ system; this cutoff scale corresponds to the ``boundary" oscillator quanta $\widetilde{N} = 2 \widetilde{n} + l $ of the 3DHO basis, which divides the Hamiltonian matrix of the $np$ system into the ``interior" region (with interaction) and the complementary ``asymptotic" region (free of interaction) as
  \begin{widetext}
\begin{gather}
\langle nlSJM_J | H | n'lSJM_J \rangle=
\begin{cases}
\  \langle nlSJM_J | T_{\rm rel} + V_{\rm NN} | n'lSJM_J \rangle & \text{for} \  n \leq \widetilde{n} \ \text{and} \ n' \leq {\widetilde{n} }; \\
\  \langle nlSJM_J | T_{\rm rel} | n'lSJM_J \rangle & \text{for} \  n > \widetilde{n} \ \text{or} \ n'> \widetilde{n} .
\end{cases} \label{eq:expanssionOftheScatteringChannel}
\end{gather}
This is the only assumption of the HORSE method. In order to improve the convergence of the scattering phase shift and radiative capture cross section, we further apply a ``smoothing" scheme \cite{Gyarmati:1979,Revai:1985} to the interaction matrix element $ \langle nlSJM_J | V_{\rm NN} | n'lSJM_J \rangle $ in Eq. \eqref{eq:expanssionOftheScatteringChannel}. In particular, we substitute $ \langle nlSJM_J | V_{\rm NN} | n'lSJM_J \rangle $ in Eq. \eqref{eq:expanssionOftheScatteringChannel} by the ``smoothed" interaction interaction matrix element
\begin{align}
\langle nlSJM_J | \widetilde{V}_{\rm NN}  | n'lSJM_J \rangle = 
\begin{cases}
\ \sigma ^n_{\widetilde{n}} \langle nlSJM_J | V_{\rm NN} | n'lSJM_J \rangle  \sigma ^{n'}_{\widetilde{n}} & \text{for} \  n \leq \widetilde{n} \ \text{and} \ n' \leq {\widetilde{n} }; \\
0 & \text{for} \  n > \widetilde{n} \ \text{or} \ n'> \widetilde{n} ,
\end{cases}
\end{align}
where the smoothing function takes the form \cite{Gyarmati:1979} 
\begin{align}
\sigma ^n_{\widetilde{n}} = \frac{1- e^{ - \big[\alpha \frac{n- (\widetilde{n}+1)}{\widetilde{n}+1} \big]^2 }  }{1-e^{-{\alpha}^2}},
\end{align}
with $\alpha $ being the dimensionless parameter. One can readily check that $\sigma ^n_{\widetilde{n}} $ has no effect for limited $\widetilde{n} $ when $\alpha \rightarrow 0 $ or $\alpha \rightarrow \infty $. In practice, one choose the $\alpha $ values that optimize convergence.

With the truncation of the interaction matrix elements, the amplitudes $\{ a_{nl}(k)\} $ are also sorted into two corresponding sets: $\{ a^{\rm int}_{nl}(k)\} $ ($n \leq {\widetilde{n}}$) and $\{ a^{\rm as}_{nl}(k)\} $ ($n > \widetilde{n}$), which are solved as follows. In the asymptotic region, the Hamiltonian is just the kinetic energy operator, which has the tridiagonal matrix form in the 3DHO representation. The amplitudes of the wave function $\{ a^{\rm as}_{nl}(k)\} $ ($n > \widetilde{n}$) obey the three-term recurrence relation:
\begin{multline}
 \langle n l S J M_J | T_{\rm rel}  | (n-1) l S J M_J \rangle a^{\rm as}_{(n-1)l}(k) +
  \langle n l S J M_J | T_{\rm rel} - E  | n l S J M_J \rangle a^{\rm as}_{nl}(k) \\
 +\langle n l S J M_J | T_{\rm rel}  | (n+1) l S J M_J \rangle a^{\rm as}_{(n+1)l}(k) = 0 , 
 \label{eq:threeTermRecurrence}
\end{multline}
where the matrix elements of the kinetic energy operator are
\begin{align}
&  \langle (n+1) l S J M_J | T_{\rm rel} | n l S J M_J \rangle = \frac{1}{2} \omega \sqrt{\Big(n+l+\frac{3}{2} \Big) \Big(n+1 \Big)} , \label{eq:KE1}  \\
& \langle n l S J M_J | T_{\rm rel} | n l S J M_J \rangle = %\frac{p_0^2}{2 \widetilde{m}_N}
\frac12\omega \Big( 2n+l+\frac{3}{2} \Big) , \label{eq:KE2}  \\
& \langle n l S J M_J | T_{\rm rel} | (n+1) l S J M_J \rangle =
\frac12\omega \sqrt{\Big(n+l+\frac{3}{2} \Big) \Big(n+1 \Big)} .
 \label{eq:KE3} 
\end{align}
We adopt two linearly independent solutions for Eq. \eqref{eq:threeTermRecurrence} (see, e.\:g., Refs.~\cite{Bang:2000bang, Yamani:1975, Zaytsev:1998,Shirokov:2003kk}):
\begin{align}
S_{nl}(k) =& (-1)^n \sqrt{ \frac{\pi r_0 n!}{v \Gamma (n+l + \frac{3}{2})} } (kr_0)^{l+1}% \nonumber \\ & \times 
\exp \Big( - \frac{k^2 r_0^2}{2} \Big) L_n^{l+\frac{1}{2}} \big( k^2 r_0^2 \big) , \\
C_{nl}(k) =&  \frac{(-1)^{n+l}}{\Gamma (-l + \frac{1}{2})} \sqrt{ \frac{\pi r_0 n!}{v \Gamma (n+l + \frac{3}{2})} } (kr_0)^{-l} \exp \Big( - \frac{k^2 r_0^2}{2} \Big) % \nonumber \\& \times 
{_1}F_{1} \Big( -n-l - \frac{1}{2}; -l + \frac{1}{2} ; k^2 r_0^2  \Big),
\end{align}
where ${_1}F_{1} (c;d;x) $ is the confluent hypergeometric function \cite{Arfken:2011}.

The asymptotic amplitudes can be expressed as a linear combination of $S_{nl}(k)$ and $C_{nl}(k)$ 
\begin{align}
a_{nl}^{\rm as}(k) = \cos \delta _l S_{nl}(k) + \sin \delta _l C_{nl}(k),
\label{assy}
\end{align}
where $\delta _l $ denotes the scattering phase shift of the partial wave with the orbital angular momentum $l$. According to Eq. \eqref{eq:threeTermRecurrence}, one notes that the above solution holds also for the case when $n = \widetilde{n}$. This will serve as the condition to match the amplitudes in the interior region with those in the asymptotic region.

The amplitudes $\{ a^{\rm int}_{nl}(k) \} $ (with $ 0 \leq n \leq \widetilde{n}$) in the interior region satisfy the algebraic equation as 
\begin{gather}
\sum _{n'=0}^{\widetilde{n}} \Big[ \langle n l S J M_J | H  | n' l S J M_J \rangle  - \delta _{nn'} E \Big] a^{\rm int}_{n'l}(k)
= -\delta _{n \widetilde{n}} \langle \widetilde{n} l S J M_J | T_{\rm rel} | (\widetilde{n}+1) l S J M_J \rangle a_{(\widetilde{n}+1)l}^{\rm as}(k) .
\end{gather}
Each amplitude $a^{\rm int}_{nl}(k) $ can be expressed in terms of $ a_{(\widetilde{n}+1)l}^{\rm as}(k) $ as \cite{Bang:2000bang}
\begin{align}
a^{\rm int}_{nl}(k) = \mathcal{G} _{n \widetilde{n} } a_{\widetilde{n} +1,l}^{\rm as}(k) , \label{eq:solnOfInteriorRegion}
\end{align}
with the matrix elements being
\begin{gather}
\mathcal{G} _{nn'} = - \sum _{\nu =0} ^{\widetilde{n}} 
\frac{ \langle nlSJM_J | \nu \rangle \langle \nu | n'lSJM_J \rangle }{E_{\nu} - E} 
\langle n'lSJM_J | T_{\rm rel} | (n'+1)lSJM_J \rangle,
\label{Gnn}
\end{gather}
where $E_{\nu }$ and $ \langle nlSJM_J |\nu \rangle $ are respectively the eigenvalue and the components of 
the corresponding eigenvector of the Hamiltonian in the interior region: 
\begin{gather}
\sum _{n'=0}^{\widetilde{n}} \  \langle nlSJM_J |  H  | n'lSJM_J \rangle 
 \langle n'lSJM_J | \nu \rangle  = E_{\nu} \langle nlSJM_J | \nu \rangle ,
 \qquad 0 \leq n \leq \widetilde{n}.
 \label{EigenEq}
\end{gather}
  \end{widetext}
The phase shift $\delta _l $ is obtained from the matching condition of the amplitudes~\eqref{eq:solnOfInteriorRegion}. In particular, one notes that the amplitude~$a^{\rm int}_{\widetilde{n} l}(k)$ satisfies Eq. \eqref{eq:threeTermRecurrence} when $n=\widetilde{n} +1$ and thus both $a^{\rm int}_{\widetilde{n} l}(k)$ and $a^{\rm as}_{\widetilde{n}+1, l}(k)$ can be expressed according to Eq.~\eqref{assy}. Therefore, using Eq.~\eqref{eq:solnOfInteriorRegion} the phase shift can be expressed as \cite{Bang:2000bang}
\begin{align}
\tan \delta _l  = -\frac{ S_{\widetilde{n} l}(k) - \mathcal{G} _{\widetilde{n} \widetilde{n} } S_{\widetilde{n} +1,l}(k) }{ C_{\widetilde{n} l}(k) - \mathcal{G} _{\widetilde{n} \widetilde{n} } C_{\widetilde{n}+1,l}(k)} .
\label{tandelta}
\end{align}

After calculating the phase shift $\delta_{l}$ at any positive energy $E$, we get the respective scattering wave function as an infinite expansion in 3DHO basis functions~\eqref{eq:scatteringWaveFunction} where at $n\geq \widetilde{n}$ the amplitudes $a_{nl}(k) =a_{nl}^{\rm as}(k)$ are calculated using Eq.~\eqref{assy} and at $n< \widetilde{n}$ the amplitudes $a_{nl}(k) =a_{nl}^{\rm int}(k)$ are calculated using Eq.~\eqref{eq:solnOfInteriorRegion}. In our calculations of the matrix elements of the $M1$ transition operator (see Subsection~\ref{sec:operator}), we restrict the sum in $n$ in Eq.~\eqref{eq:scatteringWaveFunction} by using only the 3DHO terms with oscillator quanta $2n+l\leq N_{\max}$ and verify that the accepted value of maximal allowed quanta $N_{\max}$ guarantees the convergence of the phase shift~$\delta_{l}$ and the $p(n,\gamma )d$ radiative capture cross section as well as their independence on the 3DHO basis parameter~$\omega$. We remark that we use the same value of $N_{\max}$ for the scattering wave function truncation as that for the deuteron ground state wave function (see Subsection~\ref{sec:gs}).

As has been already mentioned above, for our very low-energy application here, we have the initial scattering wave function of the $np$ system in the ${^1}S_0$ state, that is we set $l=0$, $S=0$, $J=0$, and $M_J=0$. 

\subsection{The final state\label{finalstate}\label{sec:gs}}

The final state of the $np$ system is a bound state, the deuteron, characterized by the quantum numbers
$J=1$, $S=1$ and positive parity, that means that the orbital momentum takes values $l=0$, 2. We
construct the deuteron wave function in the coordinate space as a finite expansion in the 3DHO basis functions with oscillator quanta $2n+l\leq N_{\max}$
\begin{multline}
\langle \vec{r} | \phi _f \rangle = %\sum _{n=0} ^{\infty} \sum _{l=0}^{\infty} 
\sum_{l=0,2}\sum_{n=0}^{\frac12(N_{\max}-l)} b_{nl} R_{nl} (r) \\
\times  \sum _{m_{l},m_s} (lm_lSm_s|JM_J) Y_{lm_l}(\Omega _{\hat{r}}) \chi _{Sm_s} , \label{eq:DeuteronChannel3DHO}
\end{multline}
where the amplitudes $ \{ b_{nl} \}$ satisfy a finite set of algebraic equations:
\begin{multline}
\hspace{-2.5ex}\sum_{l'=0,2}\!\sum_{n'=0}^{\frac12(N_{\max}-l')} \!
\langle nlSJ M_J | H - E \delta _{nn'} \delta _{ll'} | n'l'S J M_J \rangle b_{n'l'} = 0 , \\
l=0,2;\  n=0,1,..., \frac12(N_{\max}-l).
 \label{eq:finalStateAlgebraicEquations}
\end{multline}
The final state wave function is normalized to unity,
\begin{align}
\sum_{l=0,2}\!\sum_{n=0}^{\frac12(N_{\max}-l)} | b_{nl} |^2 =1.
\end{align}

We obtain the  amplitudes $ \{ b_{nl} \}$ by a direct  diagonalization  of the Hamiltonian  matrix $\langle n lSJ M_J | H  | n'l'S J M_J \rangle$. The truncation boundary $N_{\max}$ is chosen to be large enough and verified to provide the convergence and independence on the basis parameter $\omega$
 of the deuteron binding energy and other observables as well as of the calculated cross section of the $p(n,\gamma )d$ reaction.

The Hamiltonian of the $np$ system is degenerate in the $M_J$ values. According to the discussion in the beginning of Section~\ref{sec:TheoryPart}, we select $M_J=0$ for the final state and compute the transition probability and the radiative capture cross section according to Eqs.~\eqref{eq:Transition_Probability_Equivalence} and \eqref{eq:CaptureCrossSection}, respectively.

\subsection{Transition operator \label{sec:operator}}
The $M1$ transition in a nuclear system is facilitated by the $M1 $ operator which is defined as \cite{Bohr:1998}
\begin{align}
\label{eq:mm-op-1}
\vec{\mu} = \frac{1}{2}\int d^{3}\vec{x}\; \vec{x} \times \vec{\bar{j}}(\vec{x}) ,
\end{align}
where \(\vec{\bar{j}}(\vec{x}) \) is a nuclear electromagnetic current in coordinate space. Following convention, this operator is multiplied by a factor of $\sqrt{3/4\pi}$ when calculating the $M1$ transition \cite{Maris:2014jha}. As we are working with a two-nucleon (2N) system, we use only the operators derived from one-nucleon (1N) and 2N nuclear electromagnetic currents from Refs. \cite{Kolling:2009iq,Kolling:2011mt,Krebs:2019aka}, which are derived from the $\chi$EFT and are consistent with the LENPIC $NN$ interactions of Refs. \cite{Epelbaum:2014sza,Epelbaum:2014efa} adopted in this work. The currents in Refs. \cite{Kolling:2009iq,Kolling:2011mt,Krebs:2019aka} are in momentum space. They can be used in Eq. \eqref{eq:mm-op-1} via the Fourier transformation
\begin{align}
\label{eq:current-fourier-transformation}
\vec{\bar{j}}(\vec{x}) = \int \frac{d^{3}\vec{k}}{(2\pi)^{3}} e^{i \vec{k}\cdot\vec{x}} \vec{j}(\vec{k}),
\end{align}
where $\vec{j}(\vec{k}) $ is the momentum space current. The nuclear electromagnetic currents derived from $\chi$EFT are systematically arranged according to a power counting scheme:
\begin{align}
  \label{eq:current-order}
  \vec{j}^{a\mathrm{N}} = \vec{j}^{a\mathrm{N}}_{\mathrm{LO}} + \vec{j}^{a\mathrm{N}}_{\mathrm{NLO}} + \vec{j}^{a\mathrm{N}}_{\mathrm{N2LO}} + \dots, 
\end{align}
where the superscript $a\mathrm{N}$ indicates an $a$-nucleon current (we take $a=1$ or $2$ for the two-nucleon system). It is also worth noting that not all of the orders are present for a particular $a$-nucleon current. 

In this work, we consider only the nuclear electromagnetic currents up to N2LO in the $\chi$EFT power counting  \cite{Kolling:2009iq,Kolling:2011mt,Krebs:2019aka}. At LO, there is no contribution to the nuclear electromagnetic currents. At NLO, there are both 1N and 2N electromagnetic current operators. In particular, the 1N electromagnetic current operator at NLO is
\begin{align}
  \label{eq:single-nucleon-current-nlo}
  \vec{j}^{1\mathrm{N}}_{\mathrm{NLO}}
  % (\vec{q}_1, \vec{Q}_1, \vec{k})
  = \frac{|e|}{4m_N}
  \Bigl[
  -i [ \vec{q}_j \times \vec{\sigma}_j ] (\mu_s + \mu_v \tau_{j, z})
  + 2 \vec{Q}_j (1 + \tau_{j, z})
  \Bigr],
\end{align}
where $m_N$ denotes the nucleon mass (taken to be 938.92 MeV) and $e$ denotes the elementary charge. $\mu_{s} = 0.880 $ and $\mu_{v}= 4.706 $ are the isoscalar and isovector anomalous magnetic moments of the nucleus, respectively. \(\vec{q}_{j} = \vec{p}_{j}' - \vec{p}_{j}\), \(\vec{Q}_{j} = (\vec{p}_{j}' + \vec{p}_{j}) / 2\) are the linear combinations of the incoming (\(\vec{p}_{j}\)) and outgoing (\(\vec{p}_{j}'\)) momenta of the $j^{\rm th}$ nucleon. \(\vec{\sigma}_{j}\) denotes the spin operator of the $j^{\rm th}$ nucleon, while \(\vec{\tau}_{j}\) is the isospin operator of the $j^{\rm th}$ nucleon. The projection of \(\vec{\tau}_{j}\) is $ {\tau}_{j,z} $. Meanwhile, the 2N electromagnetic current operator at NLO is
\begin{multline}
  \label{eq:two-nucleon-current-nlo}
  \vec{j}^{2\mathrm{N}}_{\mathrm{NLO}}
      = 
      \frac{i|e|g_A^2}{4F_\pi^2}[\vec{\tau}_j \times \vec{\tau}_k]_{z}
      \frac{\vec{\sigma}_k \cdot \vec{q}_k}{q_k^2 + m_\pi^2}
      \Biggl( \vec{q}_j \frac{\vec{\sigma}_j \cdot \vec{q}_j}{q_j^2  + m_\pi^2} - \vec{\sigma}_j \Biggr) \\
        + (j \leftrightharpoons k),
\end{multline}
where $(j \leftrightharpoons k)$ indicates the term with swapped nucleon indices $j=1,2$ and $k=1,2$ (and $j \neq k$) for the two-nucleon system. $g_{A}= 1.29 $ denotes the axial coupling constant, $ F_{\pi}= 92.4 $~MeV is the pion decay constant, $ m_{\pi} = 138.03$~MeV is the average pion mass.  

At N2LO, there is a 1N current operator
\begin{align}
  \label{eq:single-nucleon-current-n2lo}
  \vec{j}^{1\mathrm{N}}_{\mathrm{N2LO}}
  = - \frac{i |e|g_{A}^{2}\tau_{j, z}}{32 \pi F_{\pi}^{2} }
  \Bigl[
  m_{\pi} - (4 m_{\pi}^{2} + q_{j}^{2}) A(|\vec{q}_{j}|)
  \Bigr]
   [\vec{q}_{j} \times \vec{\sigma}_{j}],
\end{align}
where $A({q}) = \frac{1}{2q} \tan^{-1}( \frac{q}{m_{\pi}} )  $. At this order, there is no 2N current operator.

In this work we have used a version of the LENPIC $NN$ interactions described in \cite{Epelbaum:2014efa,Epelbaum:2014sza}. These potentials have been regularized in coordinate space by multiplying them with the following coordinate space function:
\begin{align}
  \label{eq:regulator}
  f\Bigl(\frac{r}{R}\Bigr)
  & = \Biggl[1 - \exp\Biggl(-\frac{r^2}{R^2}\Biggr)\Biggr]^6,
\end{align}
where $R$ is the regulator parameter. In this work we take $R = 0.9 $ and 1.0 fm. Note that only the $V_{\rm NN}$ are regularized by $f({r}/{R}) $, while the $T_{\rm rel}$ is not regularized. Thus for consistency we also regularize the 2N current operator in Eq. \eqref{eq:two-nucleon-current-nlo}, but not the 1N current operator (in Eqs. \eqref{eq:single-nucleon-current-nlo} and \eqref{eq:single-nucleon-current-n2lo}), by multiplying it with this same function $f({r}/{R}) $. Consistency can be proved based on the continuity equation of the nuclear charge and current operators. 

We can derive the contributions from $ \vec{j}^{1\mathrm{N}}_{\mathrm{NLO}} $, $  \vec{j}^{2\mathrm{N}}_{\mathrm{NLO}} $ and $\vec{j}^{1\mathrm{N}}_{\mathrm{N2LO}} $ to the $M1 $ transition operator $\vec{\mu }$ according to Eq. \eqref{eq:mm-op-1} and Eq. \eqref{eq:current-fourier-transformation}. In particular, for the 2N system, we have\footnote{Interested readers are referred to Ref. \cite{Paletal,PaletalThesis} for detailed derivations.}
\begin{align}
\vec{\mu}^{1\mathrm{N}}_{\rm NLO} 
  = \frac{1}{2}
    \Bigl[
    (\mu_s + \mu_v \tau_{j, z}) \vec{\sigma}_{j}
    + (1 + \tau_{j, z}) \vec{L}_{j}
    \Bigr] + (j \leftrightharpoons k)  ,   \label{eq:mm-op-1n}
\end{align}
\begin{multline}
\label{eq:mm-op-2n} 
\vec{\mu}^{2\mathrm{N}}_{\rm NLO}
	= -\frac{g_{A}^{2}m_{N}m_{\pi}}{16\pi F_{\pi}^{2}}
      [\vec{\tau}_j \times \vec{\tau}_k]_{z}  
      \Bigl[
      (1 + m_{\pi}r) \big( [\vec{\sigma}_{j} \times \vec{\sigma}_{k}]\cdot\hat{r} \big) \hat{r} \\
       - m_{\pi}r [ \vec{\sigma}_{j} \times \vec{\sigma}_{k} ]
      \Bigr] 
      \frac{e^{-m_{\pi} r}}{m_{\pi} r} , 
\end{multline}
\begin{align}
\vec{\mu}^{1\mathrm{N}}_{\rm N2LO} = 0 ,      
\end{align}
where the $ \vec{\mu} ^{1\rm N}_{\rm NLO} $, $ \vec{\mu} ^{2\rm N}_{\rm NLO} $, and $\vec{\mu}^{1\mathrm{N}}_{\rm N2LO} $ operators correspond to the contributions from $ \vec{j}^{1\mathrm{N}}_{\mathrm{NLO}} $, $  \vec{j}^{2\mathrm{N}}_{\mathrm{NLO}} $, and $\vec{j}^{1\mathrm{N}}_{\mathrm{N2LO}} $, respectively. The unit vector is $\hat{r}= \vec{r}/r $.

Thus, we have the $M1$ transition operator up to N2LO as 
\begin{align}
\vec{\mu } = \vec{\mu} ^{1\rm N}_{\rm NLO} + \vec{\mu} ^{2\rm N}_{\rm NLO},
\end{align}
where both the 1N operator $\vec{\mu}^{1\mathrm{N}}_{\rm NLO} $ and the 2N operator $\vec{\mu}^{2\mathrm{N}}_{\rm NLO} $ appear at NLO according to the power counting scheme in Refs. \cite{Kolling:2009iq,Kolling:2011mt,Krebs:2019aka}. We note that, in the literature, $ \vec{\mu} ^{1\rm N}_{\rm NLO} $ and $\vec{\mu}^{2\mathrm{N}}_{\rm NLO} $ are also referred to as the impulse approximation (IA) and meson exchange current (MEC) operators, respectively. For practical numerical calculations in this work, we compute the matrix elements of $ \vec{\mu} ^{\rm 1N}_{\rm NLO}$ and $\vec{\mu} ^{\rm 2N}_{\rm NLO}$ for the $np$ system in the 3DHO representation; more details are available in Ref.~\cite{Paletal,PaletalThesis}.

\section{Results and discussion}
\label{sec:ResultAndDiscussionPart}

\subsection{Deuteron wave function and observables}

\begin{table*}[t!] 
\caption{Deuteron properties computed with the $\chi$EFT LENPIC $NN$ interactions up to N4LO with the SCS regulator $R=0.9$~fm (B) and $R=1.0$ fm (C): the ground state energy $E_{gs}$, the point-proton r.m.s. $r_{d}$, the magnetic dipole moment $\mu _D$, the electric quadrupole moment $Q$ and the $d$-wave probability $P_d$. The theoretical predictions obtained with LENPIC N4LO interactions in Refs.~\cite{Epelbaum:2014sza,Epelbaum:2014PC}, together with other empirical results \cite{VanDerLeun:1982bhg,Huber:1998zz,Mohr:2015ccw,Reid:1972re,Bishop:1979zz}, are also provided for comparison. }
\begin{ruledtabular}
\begin{threeparttable}
\begin{tabular}{c c c c c c c }
$R$ [fm] & $\chi$-Order  &  $E_{ gs}$ [MeV]                  & $r_d $ [fm] & $\mu _D$ [$\mu _N$] & $Q$ [$e \cdot {\rm fm}^2$] & $ P_d $ [$ \% $]   \\ 
\hline 
\multirow{5}{*}{$ B $} & LO    & $ -2.02347$   & 1.98975 & 0.865311 & 0.229983 & 2.5440 \\
 					& NLO    &  $-2.19867$   & 1.96828 & 0.852837 & 0.273398 & 4.7335 \\ 
				    & N2LO   &  $-2.23108$   & 1.96551 & 0.854154 & 0.270359 & 4.5025 \\
 				    & N3LO   &  $-2.22325$   & 1.97221 & 0.855946 & 0.270648 & 4.1878 \\
				    & N4LO   &  $-2.22325$   & 1.97131 & 0.855388 & 0.270985 & 4.2858 \\
\hline
\multirow{5}{*}{$ C $} & LO    & $ -2.08346$   & 1.97890 & 0.868564 & 0.214659 & 1.9731 \\ 
 					& NLO    & $-2.20609 $   & 1.96662 & 0.855659 & 0.271370 & 4.2383 \\ 
 					& N2LO   & $-2.23516 $   & 1.96436 & 0.856323 & 0.269874 & 4.1217 \\
 					& N3LO   & $-2.22326 $   & 1.97535 & 0.852620 & 0.274565 & 4.7717 \\
 					& N4LO   & $-2.22326$    & 1.97431 & 0.854718 & 0.272428 & 4.4034 \\
\hline
LENPIC-B \cite{Epelbaum:2014sza}  & N4LO & {$-2.2246$} & { 1.972} & --- & { 0.271} & {4.29} \\ 
LENPIC-B \cite{Epelbaum:2014PC}   & N4LO & {$-2.2233$}\tnote{$\ast$} & --- & --- & --- & --- \\ 
LENPIC-C \cite{Epelbaum:2014PC}   & N4LO & {$-2.2233$} & 1.9743 &  0.8547 & 0.2724 & 4.4034 \\ 
Empirical  & --- & $-2.224575(9)$ \cite{VanDerLeun:1982bhg} & 1.97535(85) \cite{Huber:1998zz} & 0.8574382311(48) \cite{Mohr:2015ccw}  & 0.2860(15) \cite{Reid:1972re,Bishop:1979zz} & ---  \\		
\end{tabular} 
        \begin{tablenotes}
            \item[$\ast$] As suggested in Ref. \cite{Epelbaum:2014PC}, relativistic corrections are necessary in order to compare with $-2.2246$~MeV in Ref. \cite{Epelbaum:2014sza}.
        \end{tablenotes}
\end{threeparttable}
\end{ruledtabular}
\label{tab:deuteronChannel}
\end{table*}

We compute the final state (deuteron) wave functions with the LENPIC $NN$ interactions up to N4LO derived with the SCS regulators $R=0.9$ or $1.0$ fm. Based on the deuteron wave functions, we calculate various deuteron properties, which include the ground state energy, the r.m.s. point charge radius, the magnetic dipole moment\footnote{Here, we make use of the one-body operator only to calculate the magnetic dipole moment.}, and the electric quadrupole moment, as well as the $d$-wave probability. In each of our calculations, a sufficiently large model space is retained for convergence analysis. We find that the results are independent of the basis parameters ($N_{\rm max}$ and $\omega$)  indicating convergence with respect to basis-space parameters. These results are presented in Table~\ref{tab:deuteronChannel}.

For each observable and choice of regulator, the expectation value also shows a converging trend as a higher order of the LENPIC $NN$ interactions is employed (the $d$-wave probability is not an observable): the order-by-order correction of the observable decreases with the chiral order of the $NN$ interaction. For comparison, we also present (1) the reference values computed by the LENPIC group with the $NN$ interactions up to the fifth order (N4LO) \cite{Epelbaum:2014sza,Epelbaum:2014PC}; and (2) respective empirical values~\cite{VanDerLeun:1982bhg,Huber:1998zz,Mohr:2015ccw,Reid:1972re,Bishop:1979zz} in Table \ref{tab:deuteronChannel}. We find that our results for the observables computed up to N4LO agree well with both independent theoretical results and with the empirical values.

The computed $d$-wave probabilities $P_d$ in the deuteron are also shown.  They should not be interpreted as the order-by-order convergence with the power expansion scheme of the LENPIC $NN$ interactions. We find that these results compare well with the corresponding results from Refs.  \cite{Epelbaum:2014sza,Epelbaum:2014PC} quoted in Table~\ref{tab:deuteronChannel}. 

We observe a moderate regulator dependence of all the computed deuteron properties: different choices of the regulator can result in a difference at the $3^{\rm rd}$ decimal place for most quantities. Exceptions are the ground state energy and the quadrupole moment results obtained with the LO $NN$ interactions, where the difference is at the $2^{\rm nd}$ decimal place.

\subsection{Scattering wave function and phase shift}

\begin{table*}[t!] 
\caption{Scattering phase shift $\delta _{0}$ (in degrees) of the $np$ system in the ${^1S_0}$ channel computed with the $\chi$EFT LENPIC $NN$ interactions up to N4LO with the SCS regulator $R=1.0$ fm for six CM bombarding energies: $E_1=1.2625 \times 10^{-8}$ MeV, $E_2 = 5 \times 10^{-7}$ MeV, $E_3 = 5 \times 10^{-4}$ MeV, $E_4 = 1 \times 10^{-3}$ MeV, $E_5 = 5 \times 10 ^{-3}$ MeV, and $E_6 =1 \times 10 ^{-2} $ MeV. The phase shifts obtained based on the LO $NN$ interaction with $R=0.9$ fm are presented in the parentheses. The phase shifts obtained by the effective range expansion (ERE) are given for comparison.
}
\begin{ruledtabular}\begin{tabular}{c c c c c c c }
 $\chi$-Order  &  $\delta _0 (E_1)  $  & $\delta _0(E_2) $  & $\delta _0(E_3) $  & $\delta _0 (E_4) $ & $\delta _0 (E_5) $ & $\delta _0 (E_6) $  \\ 
\hline 
 LO     &  0.02195 (0.02152)  &  0.1381 (0.1354)  & 4.358 (4.273)  & 6.149 (6.030)  & 13.51 (13.26)  &  18.72 (18.38)   \\
 NLO    & 0.02373   &  0.1494  &  4.711  & 6.644  & 14.56  &  20.09   \\
 N2LO   & 0.02373   &  0.1494  &  4.710  & 6.644  & 14.56  &  20.09   \\
 N3LO   & 0.02373   &  0.1494  &  4.711  & 6.644  & 14.56  &  20.09   \\
 N4LO   & 0.02373   &  0.1493  &  4.709  & 6.643  & 14.55  &  20.09   \\
\hline 
ERE     & 0.02374  & 0.1494 & 4.712 & 6.646 & 14.56 & 20.09  \\     
\end{tabular} 
\label{tab:1S0channel}
\end{ruledtabular}
\end{table*}

We calculate the initial state of the nuclear system at six CM bombarding energies: $E_1=1.2625 \times 10^{-8}$~MeV, $E_2 = 5 \times 10^{-7}$~MeV, $E_3 = 5 \times 10^{-4}$~MeV, $E_4 = 1 \times 10^{-3}$~MeV, $E_5 = 5 \times 10 ^{-3}$~MeV, and $E_6 =1 \times 10 ^{-2} $~MeV.\footnote{The $p(n,\gamma)d$ capture cross sections at these CM bombarding energies are also calculated in Ref. \cite{Rupak:1999rk} within pionless effective field theory, except for the case of $E_2$.} Working with the HORSE method, we use a sufficiently large cutoff of the boundary oscillator quanta $\widetilde{N} = 2\widetilde{n}+l$ for the interaction matrix for the $NN$ interaction in order to obtain the converged phase shift $\delta _0$. We examined that the converged phase shifts are independent on the choices of $\widetilde{N}$, and the basis parameter~$\omega $.

In Table \ref{tab:1S0channel}, we present the results of the scattering phase shift $\delta _0$ of the initial $np$ system in the ${^1S_0}$ channel as a function of (1) the LENPIC $NN$ interactions (with the SCS regulator $R=1.0$ fm); and (2) the bombarding energy $E _i$ ($i=1,\ 2, \ \dots ,\ 6$). The results based on the LENPIC $NN$ interactions with the SCS regulator $R=0.9$ fm are not shown as they agree with the results shown in Table \ref{tab:1S0channel} --- exceptions are the results based the LO $NN$ interaction (with $R=0.9$ fm), which are presented in the parentheses for comparison. 

Our results with higher-order $NN$ interactions agree well with those obtained by the effective range expansion (ERE) based on Ref. \cite{Wiringa:1995} (with the associated percentage errors evaluated to be less than $0.04\%$ of respective nominal values shown in Table \ref{tab:1S0channel}).

\subsection{Radiative capture cross section}
\begin{table*}[t!] 
\caption{The $p(n,\gamma)d$ capture cross section (in millibarns) via the $M1$ reaction channel at six CM bombarding energies (see Table~\ref{tab:1S0channel}). The theoretical predictions of the capture cross section via the $M1$ channel in Ref.~\cite{Rupak:1999rk} are also shown. The experimental cross sections of Refs.~\cite{Cox:1965,Cokinos:1977zz} and theoretical results of Refs.~\cite{Beane:2015yha,Acharya:2022} with the CM bombarding energy $E_1=1.2625\times 10^{-8}$ MeV are presented for comparison: these cross sections include also the contributions from the $E1$ reaction channel, which, however, are expected to be several orders of magnitude smaller than those from the $M1$ reaction channel \cite{Rupak:1999rk}. The bold fonts denote the chirally consistent/complete calculations up to NLO and N2LO in this work. We quote cross-section uncertainties in parenthesis based on a Bayesian analysis of chiral-order truncation of the $NN$ interaction \cite{Melendez:2019izc}. The one-sigma uncertainty is quoted for the underscored least significant digits of each result.  See the text for the discussion of the Bayesian analysis employed and other details.}
\begin{ruledtabular}
\begin{threeparttable}
\begin{tabular}{c c c c c c c c}
 & $\chi$-Order  &  $\sigma (E_1)  $ & $\sigma (E_2) $ & $\sigma (E_3) $  & $\sigma (E_4) $  & $\sigma (E_5) $ & $\sigma (E_6) $   \\ 
\hline 
\multirow{5}{*}{1N} 
					& LO     & $\underline{240}$(100) & $\underline{38}$(16)   & $1.\underline{2}$(5)   & $0.\underline{8}$(4)   & $0.\underline{36}$(15)  & $0.\underline{24}$(10) \\
 					& NLO    & $3\underline{01}$(10)  & $4\underline{7.9}$(1.6)    & $1.5\underline{0}$(5) & $1.0\underline{6}$(3) & $0.4\underline{46}$(15)   & $0.3\underline{0}$(1) \\
				    & N2LO   & $31\underline{1}$(3)   & $49.\underline{4}$(5)& $1.5\underline{50}$(15)  & $1.0\underline{9}$(1)  & $0.46\underline{0}$(5) & $0.30\underline{5}$(3) \\
 				    & N3LO   & $30\underline{7}$(3)   & $48.\underline{8}$(5)& $1.5\underline{31}$(15)  & $1.0\underline{7}$(1)  & $0.45\underline{5}$(4) & $0.30\underline{1}$(3) \\
				    & N4LO   & $30\underline{8}$(3)   & $48.\underline{9}$(5)& $1.5\underline{35}$(15)  & $1.0\underline{8}$(1)  & $0.45\underline{6}$(5) & $0.30\underline{2}$(3) \\
\hline
\multirow{5}{*}{1N$+$2N} 
					& LO     & $\underline{247}$(100) & $\underline{39}$(16)   & $1.\underline{2}$(5)   & $0.\underline{9}$(4)   & $0.\underline{37}$(15)  & $0.\underline{25}$(10) \\
 					& NLO    & {\bf 3\underline{12}(10)}  & {\bf 4\underline{9.6}(1.6)}  & {\bf 1.5\underline{6}(5)} & {\bf 1.0\underline{9}(3)} & {\bf 0.4\underline{62}(15)}  & {\bf 0.3\underline{1}(1)} \\
				    & N2LO   & {\bf 32\underline{2}(3)}  & {\bf 51.\underline{1}(5)} & {\bf 1.6\underline{05}(15)}  & {\bf 1.1\underline{3}(1)}  & {\bf 0.47\underline{7}(5)} & {\bf 0.31\underline{6}(3)} \\
 				    & N3LO   & $31\underline{9}$(3)  & $50.\underline{6}$(5)& $1.5\underline{90}$(15)  & $1.1\underline{2}$(1)  & $0.47\underline{2}$(4) & $0.31\underline{3}$(3) \\
				    & N4LO   & $31\underline{9}$(3)  & $50.\underline{8}$(5)& $1.5\underline{94}$(15)  & $1.1\underline{2}$(1)  & $0.47\underline{3}$(5) & $0.31\underline{3}$(3) \\
\hline
Ref. \cite{Rupak:1999rk} & --- & 334.2 & --- & 1.667(0) & 1.170(0) & 0.4950(0) & 0.3279(0) \\ 
Ref. \cite{Beane:2015yha} & --- & 334.9$\binom{+5.2}{-5.4}$ & --- & --- & --- & --- & --- \\
Ref. \cite{Acharya:2022} & --- & {321.0($\pm 0.7 $)} & --- & --- & --- & --- & --- \\
Expt. \cite{Cox:1965} & --- &  334.2($\pm 0.5 $) & --- & --- & --- & --- & --- \\
Expt. \cite{Cokinos:1977zz} & --- &  332.6($\pm 0.7 $) & --- & --- & --- & --- & --- \\
\end{tabular} 
\end{threeparttable}
\end{ruledtabular}
\label{tab:crossSectionR1p0}
\end{table*}

\begin{figure*}[t!] 
  \centering
  \includegraphics[width=0.4\textwidth]{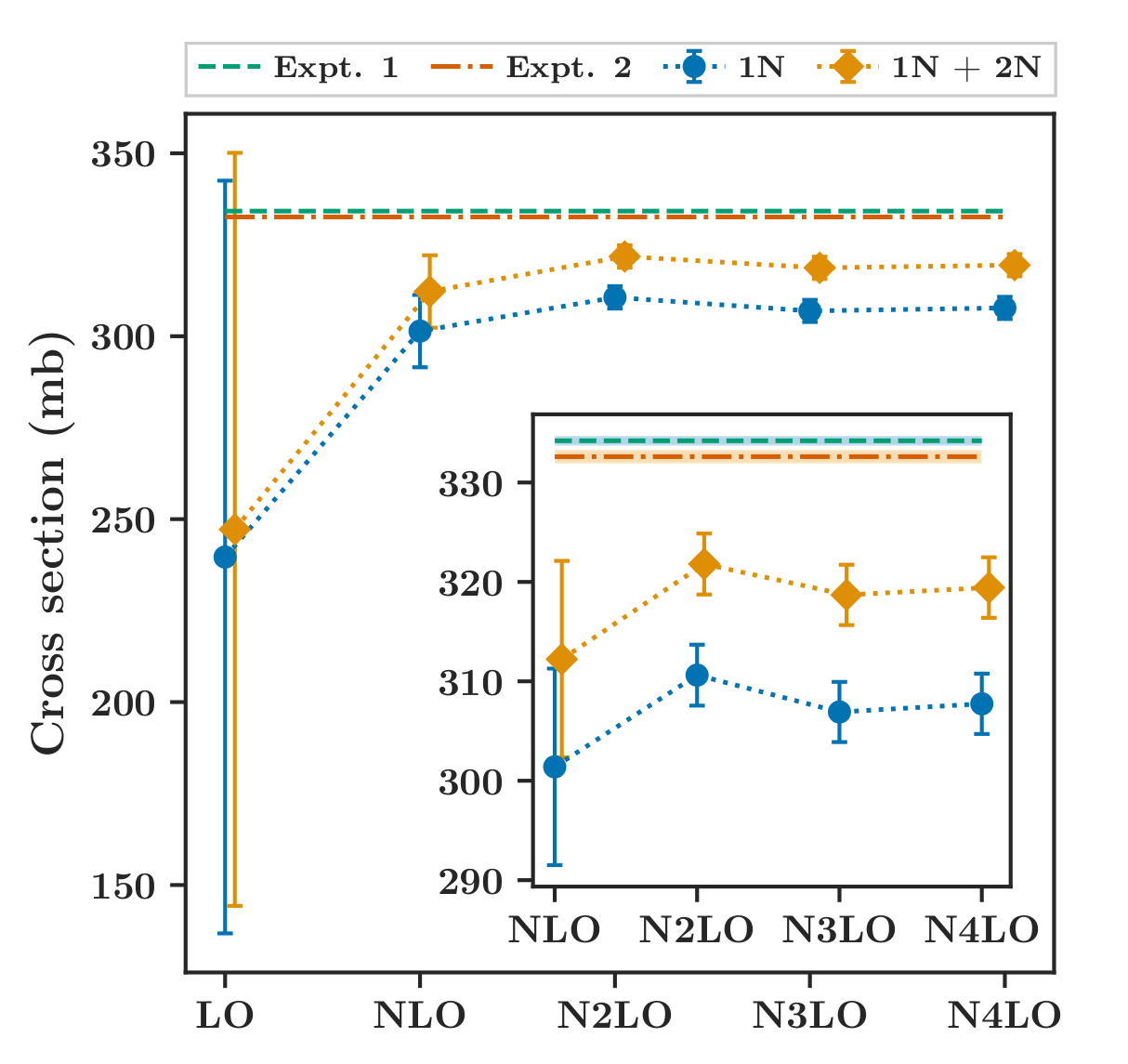}
  \caption{(color online) The $p(n,\gamma)d$ capture cross section via the $M1$ reaction channel at the CM bombarding energy \(E_1 = 1.2625 \times 10^{-8}\) MeV calculated with $\chi$EFT LENPIC $NN$ interactions up to N4LO with $R=1.0$ fm, where either 1N transition operator (blue) or both 1N$+$2N transition operators (brown) with $R=1.0$ fm are employed.  The one-sigma error bars for the cross-section results are obtained via the Bayesian analysis \cite{Melendez:2019izc} (see the text for the discussion of the Bayesian analysis employed and other details). The measured cross sections, $334.2(\pm0.5)$ mb \cite{Cox:1965} and 332.6($\pm 0.7 $) mb \cite{Cokinos:1977zz}, are also presented as green and red dashed lines (with the corresponding shaded areas denoting the error bars), respectively.} 
  \label{fig:LowestBombardingEnergyExponentialFitted}
\end{figure*}

We calculate the radiative capture cross section of the $p(n,\gamma )d$ reaction at six selected CM bombarding energies $E_1$, $E_2$, $E_3$, $E_4$, $E_5$, and $E_6$. In our calculations, we adopt the consistent SCS regulator ($R=0.9$ or $1.0$ fm) for both the transition operator and the LENPIC $NN$ interaction. In each calculation, the consistent LENPIC $NN$ interaction (i.\:e., the same chiral order and SCS regulator) is adopted to calculate both the initial and final wave functions, whereas either solely the 1N transition operator or both the 1N and 2N (1N$+$2N) transition operators are adopted. 

We study the convergence of the capture cross section at each chiral order truncation of the $\chi$EFT LENPIC $NN$ interaction with fixed choices of the transition operator (either 1N or 1N$+$2N operator). As mentioned above, in calculations of the radiative capture cross section we truncate the expansion of the scattering wave function in the series of 3DHO functions [see  Eq.~\eqref{eq:scatteringWaveFunction}] at the same oscillator quanta $2n+l\leq N_{\max}$ as in the deuteron ground state. With a sufficiently large interaction matrix truncated at $\widetilde{N}$ oscillator quanta that ensures the convergence of the phase shift of the scattering wave function, we find that the results of the capture cross section converge with $N_{\rm max}$. We also checked that the converged cross-section results are independent on $\widetilde{N}$, $N_{\rm max} $, and the 3DHO basis parameter~$\omega $. 

We also study the regulator dependence of the converged cross-section results. In particular, we find that the percentage differences between the nominal values of the cross section computed with either $R=0.9$ fm or $R=1.0$ fm regulators are: (1) $< 10 \% $ with $NN$ interactions up to LO; (2) $<1 \%$ with $NN$ interactions up to NLO; (3) $<0.6\%$ with $NN$ interactions up to N2LO; (4) $<0.01\%$ with $NN$ interactions up to N3LO; and (5) $<0.5\%$ with $NN$ interactions up to N4LO, when the 1N transition operator is employed [recall that the 1N transition operator receives no pion-current contribution and is not regularized by the regulator Eq. \eqref{eq:regulator}]. The corresponding percentage differences with the 1N$+$2N transition operator (recall the 2N part of the transition operator is regularized by the consistent SCS regulator as that of the $NN$ interaction employed) are: (1) $< 9 \% $ with LO $NN$ interactions; (2) $<0.6 \%$ with $NN$ interactions up to  NLO; (3) $<0.2\%$ with the $NN$ interactions up to N2LO, N3LO and N4LO. These differences suggest a weak regulator dependence of the cross-section results computed with higher-order $NN$ interactions.

We find that the capture cross-section results converge with the chiral expansion of the $NN$ interaction for fixed bombarding energy and transition operator (either 1N or 1N$+$2N operator), while the order-by-order corrections in the results decrease. For fixed choices of the transition operator, the one-sigma uncertainties of the cross-section results are dominated by the chiral-order-truncation uncertainties of the $NN$ interactions; these uncertainties are analyzed in the systematic framework of the Bayesian analysis \cite{Melendez:2019izc} (see Appendix for details). We adopt the viewpoint that the chiral-order uncertainty is best determined up to N2LO (where the calculations are chirally consistent with the 1N$+$2N current) and cannot be improved at higher chiral order due to the limited current we employ. Therefore, we quote chiral uncertainties at chiral orders beyond N2LO to be the same as those at N2LO.

In Table \ref{tab:crossSectionR1p0}, we present the cross-section results as functions of the chiral order of LENPIC $NN$ interactions (up to N4LO) and the transition operators (either 1N or 1N$+$2N current operators up to N2LO), where the SCS regulator regularizing both the $NN$ interactions and the transition operators is taken to be $R=1.0$ fm. Corresponding one-sigma uncertainties from the Bayesian analysis are also presented.  Within the error bars, the cross-section results computed with the SCS regulators $R=0.9$ fm agree with those obtained with $R=1.0$ fm. For the purpose of illustration, we also present the plot of the cross-section results (obtained with ${R=1.0}$~fm) for the case with the CM bombarding energy $E_1$ as a function of the chiral order of the LENPIC $NN$ interaction, and the transition operator in Fig. \ref{fig:LowestBombardingEnergyExponentialFitted}. The plots of the results with the other regulator ${R=0.9}$~fm and CM bombarding energies are similar.

Based on Table \ref{tab:crossSectionR1p0}, we find, in general, the capture cross section decreases with increasing bombarding energy for fixed $NN$ interaction and transition operator. For all the calculations, our  additional 2N transition operator enhances respective cross sections calculated with merely the 1N transition operator by a few percent. 

Based on the 1N$+$2N transition operator (recall this operator is complete up to NLO and there is no contribution to the transition operator at N2LO), we perform chirally consistent/complete calculations up to NLO and also to N2LO employing the corresponding $NN$ interactions. These chirally consistent/complete results are highlighted by the bold fonts in Table \ref{tab:crossSectionR1p0}, where the results complete up to N2LO compare well with the experiments~\cite{Cox:1965,Cokinos:1977zz}, and the theoretical predictions of Refs.~\cite{Rupak:1999rk,Beane:2015yha} based on pionless effective field theory via either perturbative or lattice QCD calculations. We remark that our chirally consistent/complete calculation up to N2LO at the bombarding energy $E_1$ compare well with the prediction in Ref. \cite{Acharya:2022}, which is computed with $\chi$EFT potential regularized by the semi-local momentum space regulator \cite{Reinert:2017usi} and the multipole expansions of the electromagnetic currents derived within $\chi$EFT \cite{Pastore:2008ui,Kolling:2009iq,Acharya:2019fij,Acharya:2020bxf}. Our chirally consistent/complete calculation up to N2LO provides an uncertainty of about $1\%$ of the nominal value, which is reasonable when compared with the N3LO chiral uncertainty of 0.2$\%$ quoted by  Acharya and Bacca \cite{Acharya:2022}.

The cross-section results with higher-order (i.e., N3LO and N4LO) $NN$ interactions are also presented in Table \ref{tab:crossSectionR1p0}. These results should be regarded as chirally incomplete (as the transition operators are only consistent up to N2LO). More systematic calculations necessitate developing the chirally consistent higher-order transition operators, which are expected to improve the precision and accuracy of our calculations \cite{Piarulli:2012bn,Acharya:2022} --- this will be the focus of future work.

\section{Summary and outlook}
\label{sec:SummaryAndOutlook}

In this work, we focus on the radiative capture of a neutron by a proton $p(n,\gamma )d$ at very low energies --- the bombarding energy in the center-of-mass frame is less or equal to $0.01$ MeV, where the $M1$ transition dominates. The input of our calculations, the $NN$ interactions and the transition operators, are from the $\chi$EFT~\cite{Weinberg:1991um,Weinberg:1990rz,Scherer:2012book}, which is a low-energy theory of quantum chromodynamics. 

In particular, we construct the Hamiltonians of the $np$ system using the $\chi$EFT LENPIC \cite{Epelbaum:2008ga} $NN$ interactions~\mbox{\cite{Epelbaum:2014efa,Epelbaum:2014sza}} up to the N4LO with the semi-local coordinate space regulators $R=0.9$ or 1.0 fm. The deuteron wave functions are obtained by a direct matrix diagonalization. These wave functions are used to compute the deuteron properties, where the results exhibit a moderate regulator dependence. We find that the computed deuteron observables converge when higher-order LENPIC $NN$ interactions are employed. Our results compare well with those of others~\cite{Epelbaum:2014sza,Epelbaum:2014PC} and with empirical results \cite{VanDerLeun:1982bhg,Huber:1998zz,Mohr:2015ccw,Reid:1972re,Bishop:1979zz}. 

We compute the scattering wave functions of the initial scattering state in the $np$ system by the HORSE method using the same $\chi$EFT LENPIC $NN$ interactions. We find that the phase shift results computed with higher-order $NN$ interactions have negligible regulator dependence and they agree well with those obtained by the effective range expansion \cite{Wiringa:1995}. 

We compute the $M1$ transition operator up to N2LO within the same $\chi$EFT framework adopted in developing the $NN$ interactions in this work. The transition operator consists of the one-nucleon (impulse approximation) and two-nucleon (meson exchange current) operators. We regularize the two-body current operator by the consistent semi-local coordinate space regulators utilized in the $NN$ interactions. 

Combining the initial and final state wave functions of the $np$ system together with the transition operator, we calculate the $p(n,\gamma )d$ reaction cross section. We find that the additional two-nucleon operator enhances the cross sections by a few percent in all calculations which improves agreement between theory and experiment where available. The regulator dependence of the cross-section results is weak when higher-order $NN$ interactions are included. Our results converge with the chiral expansion of the $NN$ interactions. The uncertainties of the cross-section results are dominated by the chiral-order truncation of the $NN$ interaction when compared with uncertainties from our numerical methods. We quantify these uncertainties by the Bayesian analysis \cite{Melendez:2019izc}.

The chirally consistent/complete calculations of the $p(n,\gamma )d$ reaction cross section are performed with the consistent $NN$ interactions and transition operator up to N2LO. The results compare well with other theoretical studies~\cite{Rupak:1999rk,Beane:2015yha,Acharya:2022} and with the experiments~\cite{Cox:1965,Cokinos:1977zz}. The calculations with $NN$ interactions of higher orders are also presented. 

Going forward, it will be important to systematically investigate the contributions from the nuclear electromagnetic current operators up to higher chiral orders. This will enable us to perform precision calculations for a wide class of photon-induced nuclear reactions. As {\it ab initio} microscopic reaction theories provide predictive power in the investigations of the radiative capture cross section (especially valuable for astrophysics applications at extremely low energies), it will also be important to generalize the current method to study the nucleon capture reactions on other nucleus. Such research will, in turn, provide an important test bed for the on- and off-shell properties of inter-nucleon interactions and insights on the nuclear response to external probes.

\section*{Acknowledgements}
We thank E. Epelbaum and R. Navarro-Perez for useful discussions and sharing numerical results from their studies. We also acknowledge fruitful discussions with P.~Maris, M. Caprio, and B. Acharya. This work was supported by Russian Foundation for Basic Research under Grant No.~20-02-00357 and by the U.S. Department of Energy under Grants No.~DESC00018223 (SciDAC/NUCLEI) and No.~DE-FG02-87ER40371. A portion of the computational resources were provided by the National Energy Research Scientific Computing Center (NERSC), which is supported by the US DOE Office of Science.

\appendix*
\section{The Bayesian analysis}
Following Ref. \cite{Melendez:2019izc}, we consider a $\chi$EFT expansion of a general scattering observable $y$ as a function of a $d$-dimensional real variable $x$:
\begin{multline}
  y(x) = y^{(0)}(x) + \Delta y^{(2)}(x) + \Delta y^{(3)}(x) + \cdots \\
       = y_{\rm ref}(x) \big(c_{0}(x) + c_{2}(x) Q^{2} + c_{3}(x) Q^{3} + \ldots \big),
\end{multline}
where $\Delta y^{(2)}(x) = y^{(2)}(x) - y^{(0)} (x)$ and $\Delta y^{(j)}(x) = y^{(j)}(x) - y^{(j - 1)}(x) $ ($j = 3,\ 4,\ \cdots$), with the superscripts denoting the chiral order. $Q$ is the ratio of the soft scale to the hard scale of the $\chi$EFT. The dimensionful quantity $y_{\rm {ref}}(x)$ sets the overall scale. The dimensionless coefficients $\{c_{0}(x),\ c_{2}(x), \ c_{3}(x), \ \cdots \} $ are assumed to be drawn from an underlying Gaussian Process with a constant mean $\bar{\mu}$, and a squared exponential kernel
\begin{align}
\kappa(x, x'; \bar{c}, h) = \bar{c}^{2}e^{-(x - x')^{T}(x - x')/(2h)^{2}},
\end{align}
with $\bar{c}$ and $h$ being the parameters. This allows us to get the analytical expressions for the posterior probability distributions of $\{ \Delta y^{(0)}(x), \ \Delta y^{(2)}(x), \ \Delta y^{(3)}(x) , \ \cdots \}$ (see Ref. \cite{Melendez:2019izc} for details). 

For the application in this work, $y$ is the cross section, and $x$ is the bombarding energy. We assume a Gaussian prior for $\bar{\mu}$, and an inverse $\chi ^{2}$ distribution for $\bar{c}$. We take the point estimates $Q = 0.31$ and $h = {0.06}$ MeV. The maximum \emph{a posteriori} values of $Q$ and $h$, that we find after carrying out the Bayesian analysis, are approximately 0.29 and $ 0.063$ MeV, respectively. The agreement with the maximum \emph{a posteriori} $Q$ and $h$ values justifies our choice of the prior $Q$ and $h$ values. Finally, we take $y_{\text{ref}}$ to be the cross-section results based on our chirally consistent and complete calculations up to N2LO.

\end{document}